\RequirePackage{lineno}
\documentclass[aps,pr,twocolumn,superscriptaddress,showpacs,preprintnumbers,amsmath,amssymb,fleqn]{revtex4}
\usepackage{color}
\usepackage[colorlinks,linkcolor=red,anchorcolor=green,citecolor=blue]{hyperref}
\usepackage{longtable}
\usepackage{graphicx} % Include figure files
\usepackage{dcolumn}  % Align table columns on decimal point
\usepackage{multirow}
\usepackage{xcolor}

\graphicspath{{ps}}

\def\Journal#1#2#3#4{{#1} {\bf #2}, #3 (#4)}

\def\PRL{ Phys. Rev. Lett.}
\def\PRD{{ Phys. Rev.} D}

\def\chicx{\chi_{cJ}}
\def\NIMA{{ Nucl. Instrum. Methods Phys. Res., Sect. A }}

\def\chic1{\chi_{c1}}
\def\chic2{\chi_{c2}}
\begin{document}

%\preprint{\vbox{
%    \hbox{Version 3.1} \\
%    \hbox{BN 1502} \\
%    \hbox{Belle Preprint 2019-07} \\
%    \hbox{KEK   Preprint 2019-05} \\
%    \hbox{hep-ex/1904.07015}}}

\title{\quad\\[0.5cm] Search for $X(3872)$ and $X(3915)$
    decay into
    $\chi_{c1} \pi^0$ in $B$ decays at Belle
}
%Search for $X(3872) \to \chi_{c1} \pi^0$ in $B$ decays
%  at Belle}

%Search for $X(3872)$ and $X(3915)$ decay into $\chi_{c1} \pi^0$ in B
%decays at Belle

%%% Paper:    X(3872) -> chi_c1 pi0 in B decays
%%% Journal:  Physical Review D (Rapid Communication)
%%% Contacts: V. Bhardwaj (vishstar@gmail.com)
%%%           S. Jia (jiasen@buaa.edu.cn)
%%%           C. Shen (shencp@buaa.edu.cn)
%%% Non-responding authors or those who said NO are commented out.
%%% ====================================================================
%%% Click the RELOAD button on your web browser to see the updated file.
%%% ====================================================================
%%% Use \input{author} to insert this material into your latex file.
%%%%% Force institutions to appear in alphabetical order when typeset.
\noaffiliation
\affiliation{University of the Basque Country UPV/EHU, 48080 Bilbao}
\affiliation{Beihang University, Beijing 100191}
%%%\affiliation{University of Bonn, 53115 Bonn}
\affiliation{Brookhaven National Laboratory, Upton, New York 11973}
\affiliation{Budker Institute of Nuclear Physics SB RAS, Novosibirsk 630090}
\affiliation{Faculty of Mathematics and Physics, Charles University, 121 16 Prague}
%%%\affiliation{Chiba University, Chiba 263-8522}
\affiliation{Chonnam National University, Kwangju 660-701}
\affiliation{University of Cincinnati, Cincinnati, Ohio 45221}
\affiliation{Deutsches Elektronen--Synchrotron, 22607 Hamburg}
%%%\affiliation{Duke University, Durham, North Carolina 27708}
\affiliation{University of Florida, Gainesville, Florida 32611}
%%%\affiliation{Department of Physics, Fu Jen Catholic University, Taipei 24205}
\affiliation{Key Laboratory of Nuclear Physics and Ion-beam Application (MOE) and Institute of Modern Physics, Fudan University, Shanghai 200443}
\affiliation{Justus-Liebig-Universit\"at Gie\ss{}en, 35392 Gie\ss{}en}
\affiliation{Gifu University, Gifu 501-1193}
\affiliation{II. Physikalisches Institut, Georg-August-Universit\"at G\"ottingen, 37073 G\"ottingen}
\affiliation{SOKENDAI (The Graduate University for Advanced Studies), Hayama 240-0193}
\affiliation{Gyeongsang National University, Chinju 660-701}
\affiliation{Hanyang University, Seoul 133-791}
\affiliation{University of Hawaii, Honolulu, Hawaii 96822}
\affiliation{High Energy Accelerator Research Organization (KEK), Tsukuba 305-0801}
\affiliation{J-PARC Branch, KEK Theory Center, High Energy Accelerator Research Organization (KEK), Tsukuba 305-0801}
\affiliation{Forschungszentrum J\"{u}lich, 52425 J\"{u}lich}
%%%\affiliation{Hiroshima Institute of Technology, Hiroshima 731-5193}
\affiliation{IKERBASQUE, Basque Foundation for Science, 48013 Bilbao}
%%%\affiliation{University of Illinois at Urbana-Champaign, Urbana, Illinois 61801}
\affiliation{Indian Institute of Science Education and Research Mohali, SAS Nagar, 140306}
\affiliation{Indian Institute of Technology Bhubaneswar, Satya Nagar 751007}
\affiliation{Indian Institute of Technology Guwahati, Assam 781039}
\affiliation{Indian Institute of Technology Hyderabad, Telangana 502285}
\affiliation{Indian Institute of Technology Madras, Chennai 600036}
\affiliation{Indiana University, Bloomington, Indiana 47408}
\affiliation{Institute of High Energy Physics, Chinese Academy of Sciences, Beijing 100049}
\affiliation{Institute of High Energy Physics, Vienna 1050}
\affiliation{Institute for High Energy Physics, Protvino 142281}
%%%\affiliation{Institute of Mathematical Sciences, Chennai 600113}
\affiliation{INFN - Sezione di Napoli, 80126 Napoli}
\affiliation{INFN - Sezione di Torino, 10125 Torino}
\affiliation{Advanced Science Research Center, Japan Atomic Energy Agency, Naka 319-1195}
\affiliation{J. Stefan Institute, 1000 Ljubljana}
%%%\affiliation{Kanagawa University, Yokohama 221-8686}
\affiliation{Institut f\"ur Experimentelle Teilchenphysik, Karlsruher Institut f\"ur Technologie, 76131 Karlsruhe}
%%%\affiliation{Kavli Institute for the Physics and Mathematics of the Universe (WPI), University of Tokyo, Kashiwa 277-8583}
\affiliation{Kennesaw State University, Kennesaw, Georgia 30144}
\affiliation{King Abdulaziz City for Science and Technology, Riyadh 11442}
%%%\affiliation{Department of Physics, Faculty of Science, King Abdulaziz University, Jeddah 21589}
\affiliation{Kitasato University, Sagamihara 252-0373}
\affiliation{Korea Institute of Science and Technology Information, Daejeon 305-806}
\affiliation{Korea University, Seoul 136-713}
\affiliation{Kyoto University, Kyoto 606-8502}
\affiliation{Kyungpook National University, Daegu 702-701}
\affiliation{LAL, Univ. Paris-Sud, CNRS/IN2P3, Universit\'{e} Paris-Saclay, Orsay}
\affiliation{\'Ecole Polytechnique F\'ed\'erale de Lausanne (EPFL), Lausanne 1015}
\affiliation{P.N. Lebedev Physical Institute of the Russian Academy of Sciences, Moscow 119991}
%%%\affiliation{Liaoning Normal University, Dalian 116029}
\affiliation{Faculty of Mathematics and Physics, University of Ljubljana, 1000 Ljubljana}
\affiliation{Ludwig Maximilians University, 80539 Munich}
\affiliation{Luther College, Decorah, Iowa 52101}
\affiliation{Malaviya National Institute of Technology Jaipur, Jaipur 302017}
\affiliation{University of Malaya, 50603 Kuala Lumpur}
\affiliation{University of Maribor, 2000 Maribor}
\affiliation{Max-Planck-Institut f\"ur Physik, 80805 M\"unchen}
\affiliation{School of Physics, University of Melbourne, Victoria 3010}
\affiliation{University of Mississippi, University, Mississippi 38677}
\affiliation{University of Miyazaki, Miyazaki 889-2192}
\affiliation{Moscow Physical Engineering Institute, Moscow 115409}
\affiliation{Moscow Institute of Physics and Technology, Moscow Region 141700}
\affiliation{Graduate School of Science, Nagoya University, Nagoya 464-8602}
\affiliation{Kobayashi-Maskawa Institute, Nagoya University, Nagoya 464-8602}
\affiliation{Universit\`{a} di Napoli Federico II, 80055 Napoli}
%%%\affiliation{Nara University of Education, Nara 630-8528}
\affiliation{Nara Women's University, Nara 630-8506}
\affiliation{National Central University, Chung-li 32054}
\affiliation{National United University, Miao Li 36003}
\affiliation{Department of Physics, National Taiwan University, Taipei 10617}
\affiliation{H. Niewodniczanski Institute of Nuclear Physics, Krakow 31-342}
\affiliation{Nippon Dental University, Niigata 951-8580}
\affiliation{Niigata University, Niigata 950-2181}
%%%\affiliation{University of Nova Gorica, 5000 Nova Gorica}
\affiliation{Novosibirsk State University, Novosibirsk 630090}
\affiliation{Osaka City University, Osaka 558-8585}
%%%\affiliation{Osaka University, Osaka 565-0871}
\affiliation{Pacific Northwest National Laboratory, Richland, Washington 99352}
\affiliation{Panjab University, Chandigarh 160014}
\affiliation{Peking University, Beijing 100871}
\affiliation{University of Pittsburgh, Pittsburgh, Pennsylvania 15260}
\affiliation{Punjab Agricultural University, Ludhiana 141004}
%%%\affiliation{Research Center for Electron Photon Science, Tohoku University, Sendai 980-8578}
%%%\affiliation{Research Center for Nuclear Physics, Osaka University, Osaka 567-0047}
\affiliation{Theoretical Research Division, Nishina Center, RIKEN, Saitama 351-0198}
%%%\affiliation{RIKEN BNL Research Center, Upton, New York 11973}
%%%\affiliation{Saga University, Saga 840-8502}
\affiliation{University of Science and Technology of China, Hefei 230026}
\affiliation{Seoul National University, Seoul 151-742}
%%%\affiliation{Shinshu University, Nagano 390-8621}
\affiliation{Showa Pharmaceutical University, Tokyo 194-8543}
\affiliation{Soongsil University, Seoul 156-743}
%%%\affiliation{University of South Carolina, Columbia, South Carolina 29208}
\affiliation{Stefan Meyer Institute for Subatomic Physics, Vienna 1090}
\affiliation{Sungkyunkwan University, Suwon 440-746}
\affiliation{School of Physics, University of Sydney, New South Wales 2006}
\affiliation{Department of Physics, Faculty of Science, University of Tabuk, Tabuk 71451}
\affiliation{Tata Institute of Fundamental Research, Mumbai 400005}
%%%\affiliation{Excellence Cluster Universe, Technische Universit\"at M\"unchen, 85748 Garching}
\affiliation{Department of Physics, Technische Universit\"at M\"unchen, 85748 Garching}
\affiliation{Toho University, Funabashi 274-8510}
%%%\affiliation{Tohoku Gakuin University, Tagajo 985-8537}
\affiliation{Department of Physics, Tohoku University, Sendai 980-8578}
\affiliation{Earthquake Research Institute, University of Tokyo, Tokyo 113-0032}
\affiliation{Department of Physics, University of Tokyo, Tokyo 113-0033}
\affiliation{Tokyo Institute of Technology, Tokyo 152-8550}
\affiliation{Tokyo Metropolitan University, Tokyo 192-0397}
%%%\affiliation{Tokyo University of Agriculture and Technology, Tokyo 184-8588}
%%%\affiliation{Utkal University, Bhubaneswar 751004}
\affiliation{Virginia Polytechnic Institute and State University, Blacksburg, Virginia 24061}
\affiliation{Wayne State University, Detroit, Michigan 48202}
\affiliation{Yamagata University, Yamagata 990-8560}
\affiliation{Yonsei University, Seoul 120-749}
% \author{A.~Abdesselam}\affiliation{Department of Physics, Faculty of Science, University of Tabuk, Tabuk 71451} % Tabuk
\author{V.~Bhardwaj}\affiliation{Indian Institute of Science Education and Research Mohali, SAS Nagar, 140306} % IISERM
\author{S.~Jia}\affiliation{Beihang University, Beijing 100191} % Beihang

\author{I.~Adachi}\affiliation{High Energy Accelerator Research Organization (KEK), Tsukuba 305-0801}\affiliation{SOKENDAI (The Graduate University for Advanced Studies), Hayama 240-0193} % KEK
% \author{K.~Adamczyk}\affiliation{H. Niewodniczanski Institute of Nuclear Physics, Krakow 31-342} % Krakow
% \author{J.~K.~Ahn}\affiliation{Korea University, Seoul 136-713} % Korea
  \author{H.~Aihara}\affiliation{Department of Physics, University of Tokyo, Tokyo 113-0033} % Tokyo
% \author{S.~Al~Said}\affiliation{Department of Physics, Faculty of Science, University of Tabuk, Tabuk 71451}\affiliation{Department of Physics, Faculty of Science, King Abdulaziz University, Jeddah 21589} % Tabuk
% \author{K.~Arinstein}\affiliation{Budker Institute of Nuclear Physics SB RAS, Novosibirsk 630090}\affiliation{Novosibirsk State University, Novosibirsk 630090} % BINP
% \author{Y.~Arita}\affiliation{Graduate School of Science, Nagoya University, Nagoya 464-8602} % Nagoya
  \author{D.~M.~Asner}\affiliation{Brookhaven National Laboratory, Upton, New York 11973} % BNL
% \author{H.~Atmacan}\affiliation{University of South Carolina, Columbia, South Carolina 29208} % SouthCarolina
% \author{V.~Aulchenko}\affiliation{Budker Institute of Nuclear Physics SB RAS, Novosibirsk 630090}\affiliation{Novosibirsk State University, Novosibirsk 630090} % BINP
  \author{T.~Aushev}\affiliation{Moscow Institute of Physics and Technology, Moscow Region 141700} % MIPT
  \author{R.~Ayad}\affiliation{Department of Physics, Faculty of Science, University of Tabuk, Tabuk 71451} % Tabuk
% \author{T.~Aziz}\affiliation{Tata Institute of Fundamental Research, Mumbai 400005} % Tata
  \author{V.~Babu}\affiliation{Tata Institute of Fundamental Research, Mumbai 400005} % Tata
  \author{I.~Badhrees}\affiliation{Department of Physics, Faculty of Science, University of Tabuk, Tabuk 71451}\affiliation{King Abdulaziz City for Science and Technology, Riyadh 11442} % Tabuk
  \author{S.~Bahinipati}\affiliation{Indian Institute of Technology Bhubaneswar, Satya Nagar 751007} % IITB
% \author{A.~M.~Bakich}\affiliation{School of Physics, University of Sydney, New South Wales 2006} % Sydney
% \author{Y.~Ban}\affiliation{Peking University, Beijing 100871} % Peking
% ...
% BINP
% \author{T.~Zivko}\affiliation{J. Stefan Institute, 1000 Ljubljana} % Ljubljana
% \author{A.~Zupanc}\affiliation{Faculty of Mathematics and Physics, University of Ljubljana, 1000 Ljubljana}\affiliation{J. Stefan Institute, 1000 Ljubljana} % Ljubljana
% \author{N.~Zwahlen}\affiliation{\'Ecole Polytechnique F\'ed\'erale de Lausanne (EPFL), Lausanne 1015} % Lausanne
\collaboration{The Belle Collaboration}

\begin{abstract}
  We report a search for $X(3872)$  and $X(3915)$ in 
  $B^+ \to \chi_{c1} \pi^0 K^+$ decays. We set an upper limit of  $\mathcal{B}(B^+ \to X(3872) K^+) \times \mathcal{B}(X(3872) \to \chi_{c1} \pi^0)$ $ < 8.1 \times 10^{-6}$  and  $\mathcal{B}(B^+ \to X(3915) K^+) \times \mathcal{B}(X(3915) \to \chi_{c1} \pi^0)$ $ < 3.8 \times 10^{-5}$  at  90\%  confidence
  level.  We also
  measure $\mathcal{B}(X(3872) \to \chi_{c1} \pi^0)/\mathcal{B}(X(3872) \to J/\psi \pi^+ \pi^-) < 0.97$
  at  90\%  confidence level.
  The results reported here are obtained from $772 \times 10^{6}$ $B\overline{B}$ events
  collected at the $\Upsilon(4S)$ resonance with the Belle detector at
  the KEKB asymmetric-energy $e^+e^-$ collider.
\end{abstract}


\maketitle

%\linenumbers

%.................
%    INTRODUCTION
%................

%%\section{Introduction}
%A decade has passed since
The $X(3872)$ state was observed for the first time by the Belle collaboration
in  2003 via its decay to $J/\psi \pi^+ \pi^-$
in the $B^+ \to J/\psi \pi^+ \pi^- K^+$
decays~\cite{Choi:2003ue}. Its mass ($3871.69 \pm 0.17$) MeV/$c^{2}$,
narrow width ($\Gamma < 1.2$ MeV)~\cite{pdg:2018}, and other
properties suggest it to be a non-conventional
$c\bar{c}$ state.
The $X(3872)$ has also been seen in other decay modes:
$D^{0} \bar{D}^{*0}$,
$J/\psi \gamma$,
$\psi(2S) \gamma$, and
$ J/\psi \pi^+ \pi^- \pi^0$
~\cite{Belle:DDstr, BaBar:Radiative, Belle:Radiative, LHCb:Radiative, BaBar:Jpsiomega}.
Very recently, a new decay mode, $\chi_{c1} \pi^0$,
was reported by BESIII~\cite{BESIII_X} in  $e^+ e^- \to \chi_{c1} \pi^0 \gamma$.
According to their measurement, 
$R^{X}_{\chi_{c1}/\psi} \equiv \mathcal{B}(X(3872) \to \chi_{c1} \pi^0)/\mathcal{B}(X(3872) \to  J/\psi \pi^+ \pi^-)$ $=$
$0.88^{+0.33}_{-0.27}\pm0.10$, where
the first uncertainty is statistical and the second is systematic.
In comparison with  conventional charmonium, this ratio seems to be large; e.g.,
$\mathcal{B}(\psi(2S) \to J/\psi \pi^0)/\mathcal{B}(\psi(2S) \to  J/\psi \pi^+ \pi^-)=3.66 \times 10^{-3}$.

If the $X(3872)$ structure is dominated by  a charmonium $\chi_{c1}(2P)$
component,  we expect the branching fraction for the pionic
transition, $X(3872) \to \chi_{c1} \pi^0$,
to be very small due to isospin breaking by the light quark masses~\cite{Volshin_0709.4474}, significantly
suppressed compared  to that for $X(3872) \to \chi_{c1} \pi^+ \pi^-$
($R \approx$ 4.0\%). The  BESIII result disfavors the
$\chi_{c1}(2P)$ interpretation of the $X(3872)$ and suggests
instead a tetraquark  or molecular state with a significant
isovector part in its wave function, which results in an enhanced single-pion
transition~\cite{Volshin_0709.4474}.

In the search for $X(3872) \to \chi_{c1} \pi^+ \pi^-$~\cite{Belle:inclusivechi},
the Belle Collaboration determined the branching fraction
$\mathcal{B}(B^+ \to X(3872) K^+) \times \mathcal{B}(X(3872) \to \chi_{c1} \pi^+ \pi^-)$ to be less than $ 1.5 \times 10^{-6}$ at 90\%
confidence level~%~\cite{CL}
(C.L.).
In addition, the Belle Collaboration observed
$B^+ \to \chi_{c1} \pi^0 K^+$ and  published the
background-subtracted $_s\mathcal{P}lot$~\cite{pivk} distribution for
$M_{\chi_{c1} \pi^0}$, which showed
no structure at the $X(3872)$ mass.
We use a similar  technique to provide a
limit on $R^{X}_{\chi_{c1}/\psi}$.

The $X(3915)$ was first observed, via its decay to $J/\psi \omega$,
by the Belle Collaboration in $B \to J/\psi \omega K$
decay~\cite{Belle:Choi:2005}. The quantum numbers of
$X(3915)$  were identified to be $J^{PC}$ = $0^{++}$~\cite{Babar:X3915}, suggesting it may be $\chi_{c0}(2P)$.
%{\color{red} However, the measured width of $20\pm5$ MeV)~\cite{pdg:2018} is
%  much smaller than the }
If $X(3915)$ is $\chi_{c0}(2P)$, its width should be
larger~\cite{Guo}. However, the measured width
($20\pm5$ MeV$/c^2$)~\cite{pdg:2018} is significantly narrower
than theoretical expectations ($>$ 100  MeV${/c^2}$).
The $J/\psi \omega$ is also expected
to be suppressed  by the Okubo-Zweig-Iizuka (OZI) rule in the $\chi_{c0}(2P)$ scenario~\cite{Olsen}.
%If $X(3915)$ is $\chi_{c0}(2P)$, then its discovery
%mode ($J/\psi \omega$) is OZI suppressed decay~\cite{Olsen}.
A $J^{PC}$ = $2^{++}$ assignment is also
consistent with our observation~\cite{Zhou_PRL}.
If $X(3915)$ is  a non-conventional $c\bar{c}$ state, then
one may expect the single pion transition to be enhanced
in $X(3915)$ decays as compared to charmonium,
where it is suppressed due to isotopic symmetry breaking.

In the study reported here, we reproduce the previous result for
$B^+ \to \chi_{c1} \pi^0 K^+$~\cite{Belle:inclusivechi,charge_conjugate},
search for the intermediate states $X$
($X$ denotes $X(3872)$ and $X(3915)$),  and measure the
product branching fraction
$\mathcal{B}(B^+ \to X K^+) \times \mathcal{B}(X \to\chi_{c1} \pi^0)$.

%  Data sample and detector
%

%\section{Data sample and detector}
We use a sample of $772 \times 10^{6}$ $B\bar{B}$ events collected
with the Belle detector~\cite{abashian} at the KEKB asymmetric-energy
$e^+e^-$ collider, operating at the $\Upsilon(4S)$ resonance~\cite{kurokawa}.
The Belle detector is a large-solid-angle spectrometer, which includes a
silicon vertex detector (SVD), a 50-layer central drift chamber (CDC),
an array of aerogel threshold Cherenkov counters (ACC), time-of-flight
scintillation counters (TOF), and an electromagnetic calorimeter (ECL)
comprised of 8736 CsI(Tl) crystals located inside a superconducting
solenoid coil that provides a 1.5~T magnetic field. An iron flux return
yoke
located outside the coil is instrumented to detect $K^{0}_{L}$ mesons and
identify muons.
The detector is described in detail elsewhere~\cite{abashian}.
Two inner detector configurations were used.
A first sample of $152 \times 10^{6}$ $B\bar{B}$ events was collected with
a 2.0-cm-radius beam pipe and a 3-layer SVD,
and the
remaining
$620 \times 10^{6}$ $B\bar{B}$ pairs were collected with a 1.5-cm-radius
beam pipe, a 4-layer SVD and a  modified CDC~\cite{SVD2_NIMA_560_1_2006}.

%%%%%%
%\section{Event selection}
We use EVTGEN~\cite{EvtGen} with QED final-state radiation by
PHOTOS~\cite{PHOTOS} for the generation of
Monte Carlo (MC) simulation events.
GEANT3-based~\cite{GEANT} MC simulation is used to model the response
of the detector
and determine the efficiency of the signal reconstruction.
Signal MC is used to estimate the efficiency and
selection criteria for reconstructing
$B^+ \to X(\to \chi_{c1} \pi^0) K^+$ decay.

We reconstruct the $B^+ \to \chi_{c1} \pi^0 K^+$ decay mode with
the same selection criteria as  those used
in  the previous analysis~\cite{Belle:inclusivechi}. To
suppress continuum background, we require the ratio of the second to
the zeroth Fox-Wolfram moment~\cite{FoxWolf} to be less than 0.5.
Charged tracks are required to originate from
the vicinity of the interaction point (IP): the distance of closest approach
to the IP
is required to be within 3.5~cm  along the beam
direction and within
1.0~cm in the plane transverse to the beam direction.
An ECL cluster is treated as a photon candidate if it is
isolated from the extrapolated charged tracks, and its energy
in the lab frame
is greater than 100~MeV.
We reject a photon candidate if the
ratio of  energy deposited in the central 3$\times$3 square of
cells
to that deposited in the enclosing 5$\times$5 square of
cells in its  ECL cluster is less than 0.85.
This helps to reduce photon candidates originating
from neutral hadrons.

 The $J/\psi$ meson is reconstructed via its decay to
$\ell^+ \ell^-$ ($\ell$ = $e$ or $\mu$) and selected by the invariant
mass  of the $\ell^+ \ell^-$ pair ($M_{\ell\ell}$).
For the dimuon mode,  $M_{\ell \ell}$ is  the
invariant mass  $M_{\mu^+\mu^-}$; for
the dielectron mode,
the four-momenta of all photons within 50 mrad cone
of the original $e^+$ or $e^-$ direction are absorbed into the
$M_{\ell \ell} \equiv M_{e^+e^-(\gamma)}$ to reduce the radiative tail.
The reconstructed invariant mass of the $J/\psi$ candidates is required
to satisfy 2.95~GeV$/c^2 < M_{e^+ e^-(\gamma)} < 3.13$~GeV$/c^2$ or
3.03~GeV$/c^2 < M_{\mu^+ \mu^-} < 3.13$~GeV$/c^2$.
For the selected $J/\psi$ candidates, a vertex-constrained fit is
applied to the charged tracks and then a mass-constrained fit
is performed to improve the momentum resolution.  The $\chi_{c1}$
candidates are reconstructed by combining a $J/\psi$ candidate with a
photon.  To reduce background from $\pi^0 \to \gamma \gamma$,
a likelihood function is employed to distinguish isolated
photons from $\pi^0$ daughters
using the  invariant mass of the  photon pair, photon energy
in the laboratory frame  and the polar angle with respect to the
beam direction in the laboratory frame~\cite{koppenburg}. 
We  combine the candidate photon with any other photon and then reject both
photons of a pair whose $\pi^0$ likelihood is larger than 0.8.
%is larger than 0.8.
For further analysis, we keep the $\chi_{c1}$ candidates
with a reconstructed invariant mass
satisfying
3.467~GeV$/c^2 < M_{J/\psi \gamma} <$~3.535 GeV$/c^2$,
which corresponds to
$[-4.5 \sigma, +2.8\sigma]$ about  the
nominal mass of the $\chi_{c1}$~\cite{pdg:2018}, where
$\sigma$ is the $\chi_{c1}$ mass resolution from the fit to the
MC simulated $J/\psi \gamma$ mass distribution.
To improve  the momentum resolution a mass-constrained fit
is applied to the selected $\chi_{c1}$ candidates.

Particle identification is performed using specific ionization
information from the CDC, time measurements from the TOF, and  the light
yield measured in the ACC. Charged kaons and pions are identified using
the $K$ likelihood ratio,
$R_{K}= \mathcal{L}_{K}/(\mathcal{L}_K + \mathcal{L}_\pi)$, where $\mathcal{L}_K$
and $\mathcal{L}_{\pi}$ are likelihood values for the kaon and pion hypotheses~\cite{pid}.  Kaon tracks are correctly identified with an
efficiency of $89.4\%$, whereas the
probability of misidentifying a pion as a
kaon is $10.1\%$ for $B^+ \to X(3872)(\to \chi_{c1} \pi^0) K^+$.

Photon pairs are kept as $\pi^0$ candidates whose invariant mass lies  in  the range 120~MeV$/c^2 < M_{\gamma\gamma} <$ 150~MeV$/c^2$
($\pm3\sigma$ about the 
nominal mass of $\pi^0$).
To reduce
combinatorial background, the $\pi^0 \to \gamma \gamma$ candidates
are also required to have an energy balance parameter
$|E_1 - E_2|/(E_1 + E_2)$ smaller than 0.8, where $E_1$ ($E_2$) is the energy
of the first (second) daughter photon
in the laboratory frame. For each
selected $\pi^0$ candidate, a mass-constrained fit is performed
to improve its momentum resolution. 
%The corrected $\pi^0$ is then combined with the
%$\chi_{c1}$ to get corrected $M_{\chi_{c1} \pi^0}$ (hereafter referred to as
%$M_{\chi_{c1} \pi^0}$, otherwise explicitly mentioned).

To identify the $B$ meson, two kinematic variables are used:
the beam-energy-constrained mass $M_{\rm bc}$ and
the energy difference  $\Delta E$.
The former is defined as
$\sqrt{E_{\rm beam}^2/c^2 - (\sum_i \vec{p}_{i})^2}/c$ and the
latter as $\sum_i E_i - E_{\rm beam}$, where $E_{\rm beam}$ is the beam
energy and $\vec{p}_{i}$  and $E_i$ are the momentum  and energy
of
the $i$-{\rm th} daughter particle in the center-of-mass (CM) frame; the
summation is over all final-state particles used
to reconstruct the $B$ candidate.
We reject candidates having $M_{\rm bc}$
less than 5.27~GeV$/c^2$ or $|\Delta E|~>$ 120~MeV.
After the reconstruction,  an average of 1.24 $B$
candidates per event is found.
When there are multiple $B$ candidates in one event,
we retain only the candidate  with the
the lowest $\chi^2$ value defined as:
\begin{equation*}
  \chi^2 = \chi^2_V + \chi^2_{\pi^0}  + (\frac{ M_{\chicx} - m_{\chicx}}{\sigma_{\chicx}})^2 + 
(\frac{ M_{\rm bc} - m_B}{\sigma_{M_{\rm bc}}})^2, 
\end{equation*}
where $\chi^2_V$  is the reduced $\chi^2$ returned by the vertex fit 
of all
charged tracks, $\chi^2_{\pi^0}$ is  the reduced $\chi^2$ for the $\pi^0$ 
mass-constrained fit, $M_{\chicx}$ is the reconstructed mass of $\chicx$, 
and
$m_{\chicx}$ and $m_B$ are the nominal masses of the $\chicx$ and $B$ mesons,
respectively. 
This method has 95\% efficiency for selecting the true candidate.

\begin{figure*}%[ht]
  \includegraphics[trim=0cm 1cm 0cm 0.25cm,height=39mm,width=53mm]{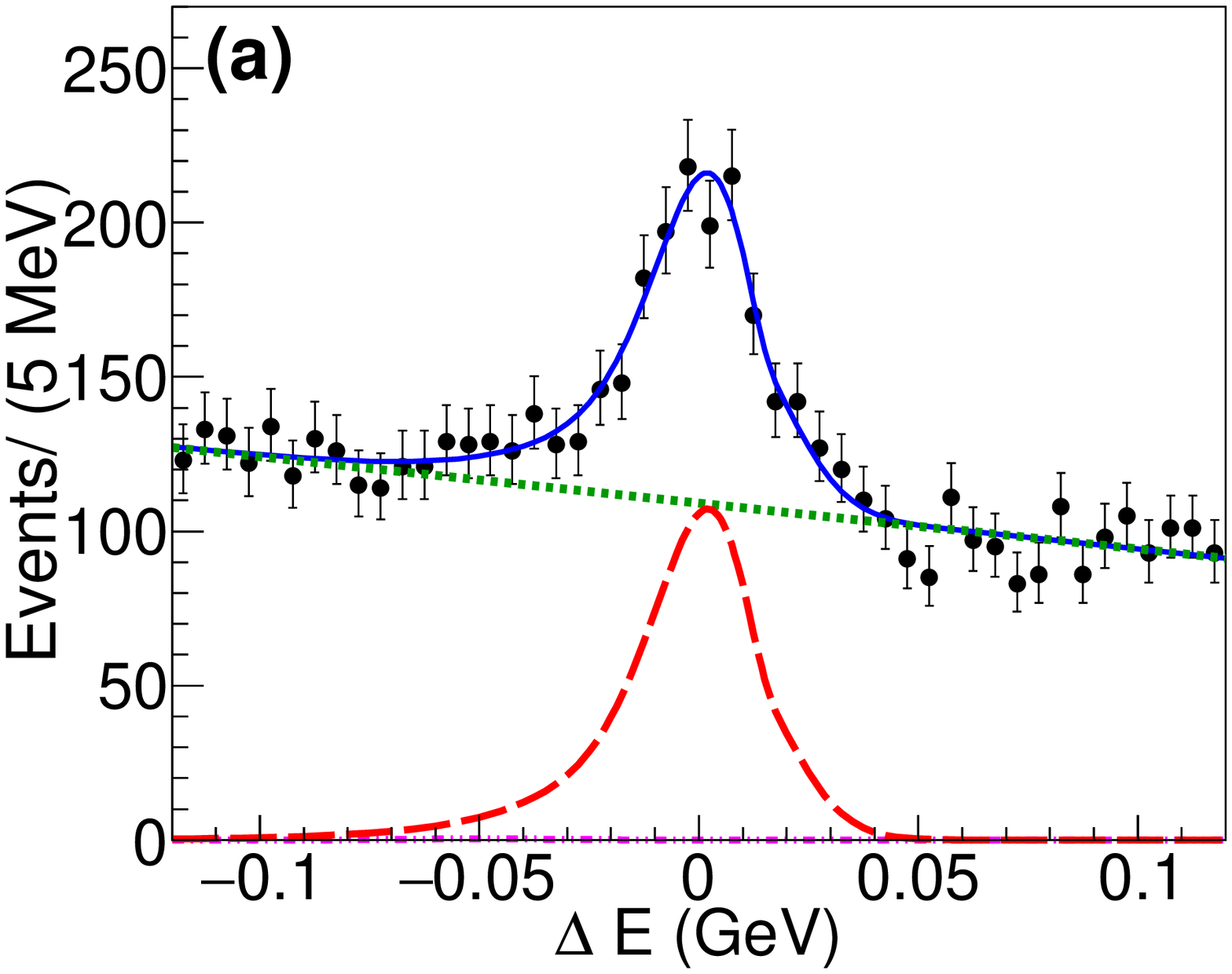}% DeltaEfit_Paper.eps}
  \includegraphics[trim=0cm 1cm 0cm 0.25cm,height=38mm,width=53mm]{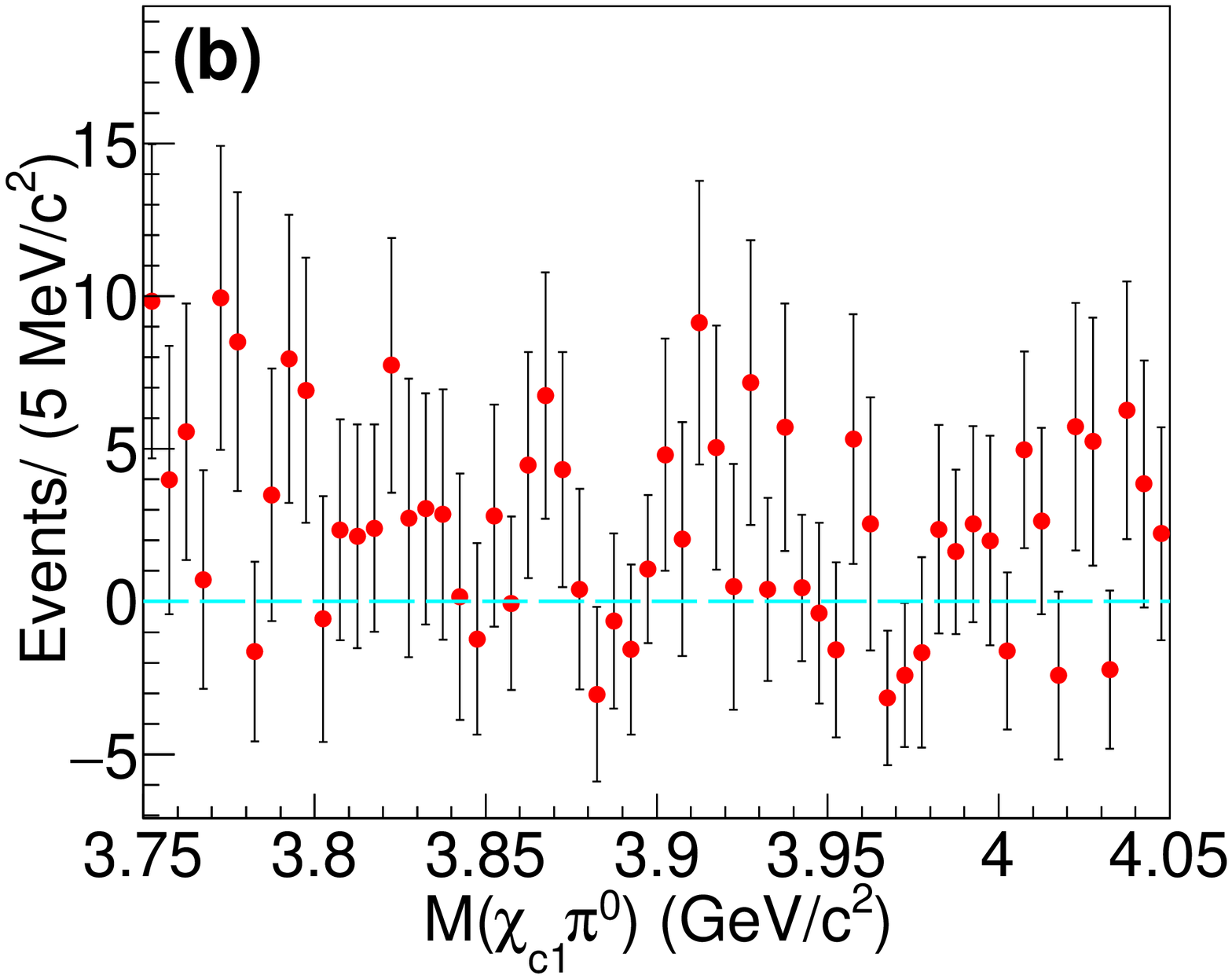}%usesPlot_Mchipi_2_5.eps}
  \includegraphics[trim=0cm 1cm 0cm 0.25cm,height=38mm,width=54mm]{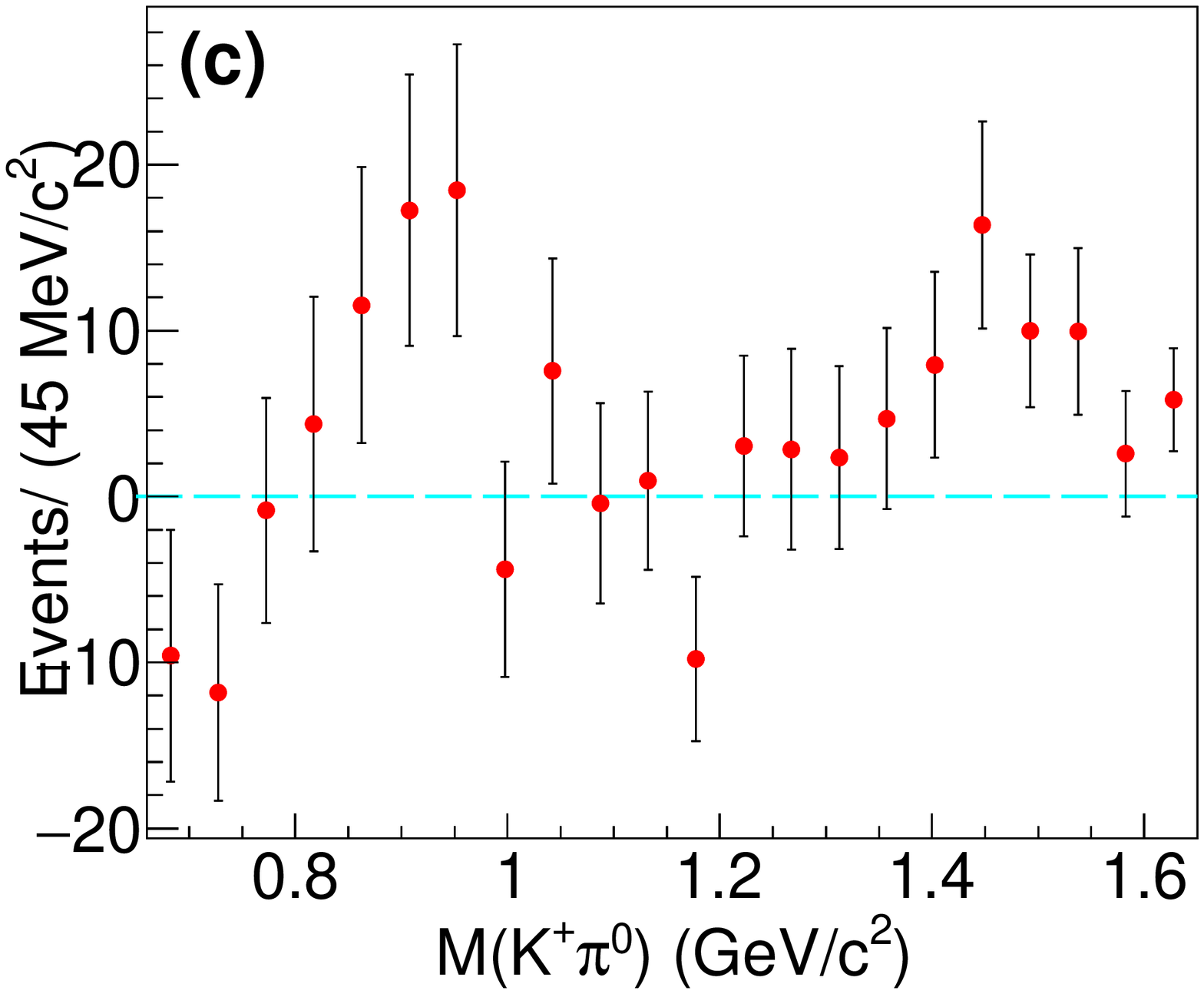}%usesPlot_MKpi_45.eps}
  \caption{ (a)  The $\Delta E$ distribution
    for the $B^0 \to \chi_{c1} \pi^0 K^+$ decay mode
    for the whole
    $M_{\chi_{c1} \pi^0}$ range.
    The curves show the signal
    (red dashed), the
    peaking background (magenta double dotted-dashed)
    and the background component (green dotted for
    combinatorial) as well as the overall fit (blue solid).
    Background-subtracted
    $_{S}\mathcal{P}$\textit{lot}
    (b) $M_{\chi_{c1} \pi^0}$ and  (c) $M_{K ^+\pi^0}$  distributions
    (in 3.75 GeV$/c^2<$
    $M_{\chi_{c1} \pi^0} <$ 4.05 GeV$/c^2$ signal window) for the
    $B^+ \to \chi_{c1} \pi^0 K^+$ decay mode. 
    Points with error bar represent the data.
  }
  \label{fig:B_chicJ_Kpi}
\end{figure*}
%\end{multicols}

%\end{multicols}
We extract the signal yield from an unbinned
extended maximum likelihood (UML) fit to the $\Delta E$ distribution.
The signal probability density function (PDF)
is modeled by a sum
of a Gaussian function and
a logarithmic Gaussian function~\cite{lg}.
The mean and width of the core
Gaussian with larger fraction are floated and the remaining parameters of tail
distribution are fixed  from studies of MC simulation.

To study the background  from events with a $J/\psi$, we use 
MC-simulated
$B \to J/\psi X$ sample corresponding to 100 times
the integrated luminosity of the data sample.
Possible peaking backgrounds from the
feed-across
of $B^+ \to \chi_{c2} \pi^0 K^+$ are found in the
$\Delta E$ distribution around $-$50 MeV, which are due
to the mass-constrained fit
to $\chi_{c1} \to J/\psi \gamma$ candidates; we estimate that
only five such events
are expected in real data.
Thus, we fix this peaking background contribution in the fit.
%for remainder of the analysis.
The PDF for the peaking background is modeled by an asymmetric
Gaussian distribution
for which the  parameters are fixed according to MC simulation after
MC/data correction (using the signal events whose mean and sigma
of the core Gaussian are floated).

The rest of the background is combinatorial and modeled by using a first-order Chebyshev polynomial.
The fit to the $\Delta E$ distribution for
$B^+ \to \chicx  \pi^0 K^+$ is shown in Fig.~\ref{fig:B_chicJ_Kpi}(a).
We obtain $806\pm 69$
signal events for the $B^+ \to \chi_{c1} \pi^0 K^+$ decay mode, 
which is consistent
with our previous study~\cite{Belle:inclusivechi}.
In order to improve the resolution on
the invariant mass of the combined $\chi_{c1}$ and $\pi^{0}$
candidates ($M_{\chi_{c1} \pi^0}$), we scale the energy and momentum of the
$\pi^0$, such that $\Delta E$ (defined below) is equal to zero while
the $M_{\pi^0}$ is  kept constant to its already mass-constrained value.
This  corrects for  the  incomplete energy measurement of the
$\pi^0$ detection. %measurement in the ECL.
The corrected four-momentum of the
$\pi^0$  is then used to improve the invariant mass $M_{\chi_{c1} \pi^0}$ and
$M_{K^+ \pi^0}$.

To search for the
$X$, we  examined the
background-subtracted $M_{\chi_{c1}\pi^0}$ distribution
produced   with the
$_{S}\mathcal{P}$\textit{lot} technique~\cite{splot} for the range
(3.75~GeV$/c^2 ~<~ M_{\chi_{c1} \pi^0}~ <$~4.05~GeV$/c^2$) as shown in
Fig.~\ref{fig:B_chicJ_Kpi}(b).  Figure~\ref{fig:B_chicJ_Kpi}(c) shows the
$M_{K \pi^0}$
$_{S}\mathcal{P}$\textit{lot} distribution in the range of interest
(3.75 GeV$/c^2 ~<~ M_{\chi_{c1} \pi^0}~ <$ 4.05 GeV$/c^2$),
 where most events come from the
$K^*$ decays. 

In order to extract the $X$ signal yield, we use the $M_{\chi_{c1} \pi^0}$
distribution within the signal-enhanced window of
$-30$ MeV $< \Delta E <$ 20 MeV for $B^+ \to (\chi_{c1} \pi^0) K^+$
candidates.
We veto events from $B^+ \to \chi_{c1} K^{*+}$  decay by rejecting
events with 791.8 MeV/$c^2~<~M(K^+\pi^0)~<$ 991.8 MeV$/c^2$.
This requirement reduces the background by 32\% with a
signal efficiency of 84\%.
We extract the signal by
performing a 1D UML fit to the $M_{\chi_{c1} \pi^0}$ distribution. The signal
PDFs for both $X(3872)$ and $X(3915)$ are modeled by the sum of two Gaussians.
All the PDF parameters are fixed from the MC simulation
after  a MC/data correction estimated from the
$B^+ \to \psi(2S)(\to \chi_{c1} \gamma) K^+$ sample is applied~\cite{Belle:chic1gamma} (the mean and sigma of the core Gaussian were fixed after scaling, while the tail parameters were fixed from signal MC).

\begin{figure}%[h!]
  \centering
  \includegraphics[trim=0.1cm 0.5cm 0.25cm 0.25cm,height=50mm,width=85mm]{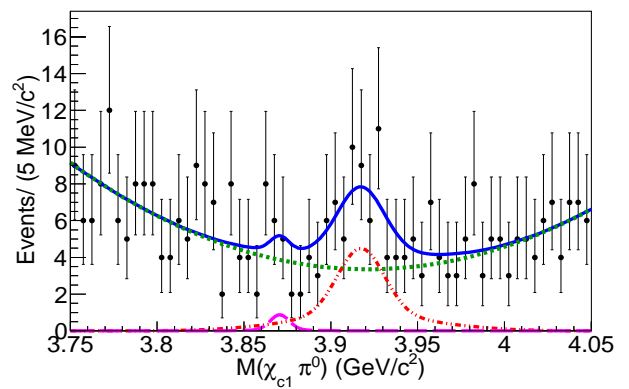}

  \caption{ 1D UML fit to the $M_{\chi_{c1} \pi^0}$ distribution
    in the $-30$ MeV $< \Delta E < $ $20$ MeV signal region for the
    $B^+ \to (\chi_{c1} \pi^0) K^+$ decay mode.
    The curves show the $B^+ \to X(3872) (\to \chi_{c1} \pi^0) K^+$ signal
    (magenta dashed), $B^+ \to X(3915) (\to \chi_{c1} \pi^0) K^+$ signal
    (red double dotted-dashed), and the background component (green dotted for
    combinatorial) as well as the overall fit (blue solid).
    Points with error bar represent the data.
  }
  \label{fig:X3872search}
\end{figure}

The efficiency ($\epsilon$) is estimated to be $5.35 \%$  and $5.37\%$ for
$B^+ \to X(3872) (\to \chi_{c1} \pi^0) K^+$  and
$B^+ \to X(3915) (\to \chi_{c1} \pi^0) K^+$ using the MC simulations,
respectively. 
This efficiency
has been calibrated by the difference between MC simulation and data,
as described later. A fit  to the data shown in Fig.~\ref{fig:X3872search}
results in a signal yield of $2.7 \pm 5.5$  ($42\pm14$)
events having significance of 0.3 $\sigma$ (2.3 $\sigma$)
for the $B^+ \to X(3872) (\to \chi_{c1} \pi^0) K^+$
($B^+ \to X(3915) (\to \chi_{c1} \pi^0) K^+$) decay mode.
The systematic uncertainty (explained later) has been included in the
significance calculation.

With the absence of any significant signal, we estimate an upper limit (U.L.) at
90\% C.L.
%We determine the U.L. by using
We apply a frequentist method that uses ensembles of
pseudoexperiments. For a given signal yield, sets of signal and
background events are generated according to their PDFs and fits are
performed. The C.L. is determined from the fraction of samples that give a
yield larger than that of data.
We estimate the branching fraction according to the formula
$\mathcal{B} = Y^{U.L.} /(\epsilon \times  \mathcal{B}_s \times N_{B\bar{B}})$;
here $Y^{U.L.}$ is the estimated U.L. yield at $90\%$ C.L.,
$\epsilon$ is the reconstruction efficiency,
$\mathcal{B}_s$ is the product of
secondary branching fraction taken from
Ref.~\cite{pdg:2018}, and $N_{B\bar{B}}$ is the number of $B \bar{B}$ mesons
in the data sample. Equal production of neutral and
charged $B$ meson pairs in the $\Upsilon(4S)$ decay is assumed.
For this assumption, an uncertainty of 1.2\% is added to
the total systematics.

We estimate the U.L. on the product  of  branching fractions
$\mathcal{B}(B^{+}\to X(3872)K^{+})\times\mathcal{B}(X(3872)\to\chi_{c1}\pi^0)$
directly from the above MC pseudoexperiment samples. The limit includes the
systematic uncertainties from efficiency,
particle identification, and signal extraction method
into the yield obtained by smearing the assumed
values by their uncertainties.
Along with that we also smear the $N_{B\bar{B}}$ and
secondary branching fraction  by adding their systematic
uncertainties as a fluctuation of the
value used to calculate the branching fraction. Using the
MC pseudoexperiment samples we
estimate the U.L. (90\% C.L.) on the product 
branching fraction as:

\begin{eqnarray*}
  \mathcal{B}(B^{+}\to X(3872)K^{+})\times\mathcal{B}(X(3872)\to\chi_{c1}\pi^0)< 8.1 \times 10^{-6} \\
  \mathcal{B}(B^{+}\to X(3915)K^{+})\times\mathcal{B}(X(3915)\to\chi_{c1}\pi^0)<3.8 \times 10^{-5}.
\end{eqnarray*}

To measure the  $R^{X}_{\chi_{c1}/\psi}$, we use the previous Belle
measurement of
$\mathcal{B}(B^+ \to X(3872) K^+) \times \mathcal{B}(X(3872) \to J/\psi \pi^+ \pi^-)$ = $(8.63\pm 0.82({\rm stat.}) \pm 0.52({\rm syst.})) \times 10^{-6}$~\cite{Belle:Choi:2011}.
Some of the systematic uncertainties cancel, such as lepton identification,
$\mathcal{B}(J/\psi \to \ell \ell)$, some  tracking systematics, and
kaon identification.
The U.L. on $R^{X}_{\chi_{c1}/\psi}$ is estimated in the same manner
as  that on
$\mathcal{B}(B^{+}\to X(3872)K^{+})\times\mathcal{B}(X(3872)\to\chi_{c1}\pi^0)$.
We remove the cancelled systematic uncertainties
and smear the pseudoexperiments with the remaining ones.
We further smear $\mathcal{B}(B^+ \to X(3872) K^+) \times \mathcal{B}(X(3872) \to J/\psi \pi^+ \pi^-)$  by  its statistical uncertainty and
uncancelled systematic uncertainties. For each toy  sample, $R^{X}_{\chi_{c1}/\psi}$
is estimated for
the generated $R^{X}_{\chi_{c1}/\psi}$. The C.L. value is  then
determined from the fraction of samples of  pseudoexperiments
having $R^{X}_{\chi_{c1}/\psi}$ larger than the central value of data.
We estimate the U.L. to be $R^{X}_{\chi_{c1}/\psi}$ $<$ 0.97 at $90\%$ C.L.

\begin{table}[h]
  \caption{ Summary of the systematics uncertainties for the
    $\mathcal{B}(B^+ \to X K^+) \times \mathcal{B}(X \to \chi_{c1} \pi^0)$ and  $R^X_{\chi_{c1}/\psi}$. }
  \begin{center}
    \begin{tabular}{lcccc}
      \hline \hline
    %  Source & \multicolumn{2}{c}{Systematic} (\%) \\ \hline
      Source & \multicolumn{2}{c}{$\mathcal{B}$ (\%)} & $R^X_{\chi_{c1}/\psi}$ (\%)  \\ %\hline
      & $X(3915)$ & $X(3872)$ & \\ \hline
      Lepton identification & 2.3 & 2.2 & - \\ %\hline
      Kaon identification & 1.0 & 1.0 & - \\ %\hline
      Efficiency & 0.5 & 0.5  & 2.2 \\ %\hline
      $B\bar{B}$ pairs & 1.4 & 1.4   &  - \\ %\hline
     $B$ production & 1.2 & 1.2 & - \\
      Tracking & 1.1 & 1.1 & 0.7 \\ %\hline
      $\gamma$ identification & 2.0 & 2.0 & 2.0 \\ %\hline
      $\pi^0$ veto & 1.2 & 1.2 & 1.2 \\ %\hline
      $\pi^0$ reconstruction & 2.2 & 2.2 & 2.2 \\ %\hline
      Signal extraction & $^{+16.1}_{-19.5}$ &  $^{+37.0}_{-44.4}$ & $^{+37.1}_{-44.5}$ \\ %\hline
      Secondary $\mathcal{B}$ & 3.0 & 3.0 & 2.9\\ \hline

      Total &  $^{+17.0}_{-20.2}$ &  $^{+37.4}_{-44.7}$  & $^{+37.4}_{-44.8}$ \\ \hline \hline
    \end{tabular}
    \label{tab:systerrall}
  \end{center}
\end{table}

Table~\ref{tab:systerrall} summarizes  systematic uncertainties for the measured product branching fraction
$\mathcal{B}(B^+ \to X K^+) \times \mathcal{B}(X \to \chi_{c1} \pi^0)$ and the ratio $R^{X}_{\chi_{c1}/\psi}$.
A correction for the small difference in the signal detection efficiency between MC
and data is applied for the lepton identification requirements,
which are determined from 
$e^+ e^- \to e^+ e^- \ell^+ \ell^-$ and $J/\psi \to \ell^+ \ell^-$
($\ell = e$ or $\mu$) samples.  Dedicated $D^{*+} \to D^0(K^- \pi^+)\pi^+$
samples are used to estimate the kaon (pion) identification
efficiency correction.  The uncertainty on the efficiency
due to
limited MC statistics is 0.5\%, and the uncertainty on the number of
$B\bar{B}$ pairs is 1.4\%. The uncertainty on the track finding efficiency is
found to be 0.35\% per track by comparing data and MC for $D^* \to D^0 \pi$
decay, where $D^0 \to K^0_S \pi^+ \pi^-$ and $K_S^0 \to \pi^+ \pi^-$.
The uncertainty on the photon identification is estimated to be 2.0\%
from a sample
of radiative Bhabha events. The systematic uncertainty associated with the
difference of the $\pi^0$ veto between data and MC is estimated to be 1.2\% from
a study of the $B^+ \to \chi_{c1}(\to J/\psi \gamma)K^+$ sample.
For $\pi^0$ reconstruction, the efficiency correction  and systematic uncertainty are estimated
from a sample of $\tau^- \to \pi^- \pi^0 \nu_\tau$ decays. The errors on
the PDF shapes are obtained by varying all fixed parameters by $\pm 1 \sigma$
and taking the change in the yield as the systematic uncertainty.
The largest uncertainty in the PDF parameterization
for $X(3872)$ ($X(3915)$)  is  $30\%$  ($^{+15}_{-17}\%$)
from fixing the  mass (width)  of the $X(3872)$ ($X(3915)$) to
the value reported in Ref.~\cite{pdg:2018}. In order to estimate the
uncertainty  coming from the
background shape, we used a third-order polynomial and took the
difference as the uncertainty. Further,  we also used large fitting
range and added the difference in quadrature to the uncertainty coming from signal extraction
procedure.
The uncertainties due to the secondary branching fractions are also taken into
account. Assuming all the sources are independent we add them in
  quadrature to obtain the  total systematic uncertainties.

To summarize, in our searches for $X(3872)$ and $X(3915)$
decaying to $\chi_{c1} \pi^0$, we did not
find a  significant signal. We obtained $2.7\pm5.5$
($42 \pm14$)  events, with a signal significance of
  0.3 $\sigma$ (2.3 $\sigma$) for the $B^+ \to X(3872) (\to \chi_{c1} \pi^0) K^+$
($B^+ \to X(3915) (\to \chi_{c1} \pi^0) K^+$) decay mode.
We determine an U.L. on the product  branching fractions
$\mathcal{B}(B^+ \to X(3872) K^+) \times \mathcal{B} (X(3872) \to \chi_{c1} \pi^0)$
$<$  $8.1 \times 10^{-6}$  and $\mathcal{B}(B^+ \to X(3915) K^+) \times  \mathcal{B}(X(3915) \to \chi_{c1} \pi^0)$
$<$  $3.8 \times 10^{-5}$  at 90\% C.L.
The  null result for our search is compatible with the interpretation of
$X(3872)$ as an admixture state of a
$D^0 \bar{D}^{*0}$ molecule and a $\chi_{c1}(2P)$ charmonium state~\cite{Volshin_0709.4474}.
One can further estimate $R^{X}_{\chi_{c1}/\psi}$ $<$ 0.97 at 90\% C.L.
Our U.L. does not contradict the BESIII result~\cite{BESIII_X}.
%However, this suggests that
%$R^{X}_{\chi_{c1}/\psi}$ is closer to the lower side of what
%BESIII observed.
This information can be used to
constrain the tetraquark/molecular component of the $X$ states.

%\acknowledgments

%\section{Acknowledgments}
We thank the KEKB group for the excellent operation of the
accelerator; the KEK cryogenics group for the efficient
operation of the solenoid; and the KEK computer group, and the Pacific Northwest National
Laboratory (PNNL) Environmental Molecular Sciences Laboratory (EMSL)
computing group for strong computing support; and the National
Institute of Informatics, and Science Information NETwork 5 (SINET5) for
valuable network support.  We acknowledge support from
the Ministry of Education, Culture, Sports, Science, and
Technology (MEXT) of Japan, the Japan Society for the 
Promotion of Science (JSPS), and the Tau-Lepton Physics 
Research Center of Nagoya University; 
the Australian Research Council including grants
DP180102629, % Sevior
DP170102389, % Varvell
DP170102204, % Yabsley
DP150103061, % Urquijo
FT130100303; % Urquijo;
Austrian Science Fund (FWF);
the National Natural Science Foundation of China under Contracts
No.~11435013,  %Zhen-An Liu
No.~11475187,  %Chang-Zheng Yuan
No.~11521505,  %Chang-Zheng Yuan
No.~11575017,  %Cheng-Ping Shen
No.~11675166,  %Wen-Biao Yan
No.~11705209;  %Yi-Ming Li
Key Research Program of Frontier Sciences, Chinese Academy of Sciences (CAS), Grant No.~QYZDJ-SSW-SLH011; % Chang-Zheng Yuan
the  CAS Center for Excellence in Particle Physics (CCEPP); %Chang-Zheng Yuan, …
the Shanghai Pujiang Program under Grant No.~18PJ1401000;  %Tao Luo
the Ministry of Education, Youth and Sports of the Czech
Republic under Contract No.~LTT17020;
the Carl Zeiss Foundation, the Deutsche Forschungsgemeinschaft, the
Excellence Cluster Universe, and the VolkswagenStiftung;
the Department of Science and Technology of India; 
the Istituto Nazionale di Fisica Nucleare of Italy; 
National Research Foundation (NRF) of Korea Grants
No.~2015H1A2A1033649, No.~2016R1D1A1B01010135, No.~2016K1A3A7A09005
603, No.~2016R1D1A1B02012900, No.~2018R1A2B3003 643,
No.~2018R1A6A1A06024970, No.~2018R1D1 A1B07047294; Radiation Science Research Institute, Foreign Large-size Research Facility Application Supporting project, the Global Science Experimental Data Hub Center of the Korea Institute of Science and Technology Information and KREONET/GLORIAD;
the Polish Ministry of Science and Higher Education and 
the National Science Center;
the Grant of the Russian Federation Government, Agreement No.~14.W03.31.0026; % from 15.02.2018;
the Slovenian Research Agency;
Ikerbasque, Basque Foundation for Science, Spain;
the Swiss National Science Foundation; 
the Ministry of Education and the Ministry of Science and Technology of Taiwan;
and the United States Department of Energy and the National Science Foundation.


\begin{thebibliography}{99}
\bibitem{charge_conjugate}
  Charge-conjugate decays are included unless explicitly
  stated otherwise.
\bibitem{lg}
The logarithmic Gaussian is parameterized as:\\
$f(x)=N_0/c(x) \times$ $\exp{- \{\ln[(\epsilon-x)/(\epsilon-x_p)]\}^2/(2\sigma^2_0)}$
%$\exp{- {\ln[(\epsilon - x)/(\epsilon - x_p)]}^2}/(2\sigma_0^2)}$,
where $\epsilon = \sigma/a + x_p$, $c(x)=\sqrt{2\pi}\sigma_0(\epsilon -x)$ and
$\sigma_0 = (\ln(a\sqrt{2\ln 2} + \sqrt{1 + 2a^2\ln 2})/\sqrt{2\ln 2}$.
Here, $N_0$ is the  normalization, $\sigma$ is the standard deviation,
$x_p$ is the mean and $a$ is the asymmetry.
\end{thebibliography}

\end{document}